\documentclass[11pt]{article}
\usepackage{acl2015}
\usepackage{times}
\usepackage{url}
\usepackage{latexsym}
\usepackage{multirow}
\usepackage{caption}
\usepackage{underscore}

\title{Given Users Recommendations Based on Reviews on Yelp }

\author{
  Shuwei Zhang \\
  {\tt shuweiz@usc.edu} \\\And
  Maiqi Tang \\
  {\tt maiqitan@usc.edu} \\\And
  Qingyang Zhang \\
  {\tt qzhang56@usc.edu} \\\AND
   Yucan Luo \\
  {\tt yucanluo@usc.edu} \\\And
  Yuhui Zou \\
  {\tt yuhuizou@usc.edu}
  }

\date{}

\begin{document}

\maketitle

\section{Introduction}
In our project, we focus on NLP based hybrid recommendation systems. Our data is from Yelp Data. For our hybrid recommendation system, we have two major components: the first part is to embed the reviews with Bert model and word2vec model; the second part is the implementation an item-based collaborative filtering algorithm to compute similarity of each review under different categories of restaurants. At the end, with the help of similarity scores, we are able to recommend users the most matched restaurant based on their recorded reviews. 

The coding work is split to several parts: selecting samples and data cleaning, proprecessing, embedding, computing similarity, and computing prediction and error. Due to the size of the data, each part will generate one or more json files as the milestone to reduce the pressure to memory and the communication between each part. 

\section{Methods}

The first step is selecting valid samples from the data sets. We found that the restaurant and users which only received and made a few reviews, and the reviews with only a few words provided limited resource for the NLP based hybrid recommendation system. A valid sample of data can be selected by setting some thresholds based on the above condition. The second step is to prepare the review data with data cleaning and preprocessing. We also want to see how the different types of preprocessing and embedding affect the result, so we generated two data sets  after the preprocessing step: one with lemmatization, and one without lemmatization. Then the data sets are split into train and test data.

In the first step, we applied four embedding models for comparison: word2vec-google-news-300 model, training a word2vec model, glove-wiki-gigaword-300 model and Bert model. With the embeded vectors, we then calculate the similarity between two groups of restaurant reviews. Finally, we used the similarity and ratings from other users to compute the prediction ratings and errors of the improved algorithm, and compared it to the result of the original item-filtering algorithm to show the improvement.

In the second step, we compared the original item CF and item CF with reviews to demonstrate the advantage our our system. For original item CF, the main goal is to match users’ rated items to other similar items and then use the ratings of other similar items to predict the rating of current user rated item. 

\begin{tabular}{ |p{1cm}|p{1cm}|p{1cm}|p{1cm}|p{1cm}|  }
 \hline
 \multicolumn{5}{|c|}{Users Ratings Over Items} \\
 \hline
& Item 1 & Item 2 & Item 3 & Item 4\\
 \hline
User 1 & 1 & 3 & N/A & 4\\
\hline
User 2 & 4 & ? & 2 & 3\\
\hline
User 3 & N/A & 5 & 5 & 4 \\
\hline
User 4 & 5 & 2 & 5 & N/A \\
 \hline
\end{tabular}
\captionof{table}{Example data set}

    For example, if we have a data set as the table 1 above, and we want to predict the rating of Item 2 rated by User 2. We use the following weight function to calculate the weight of each pair of items. 

$$w_{i, j} = \frac{\sum_{u \in U}(r_{u, i} - \bar{r_i})(r_{u, j} - \bar{r_j})}{\sqrt{\sum_{u \in U}(r_{u, i} - \bar{r_i})^2}\sqrt{\sum_{u \in U}(r_{u, j} - \bar{r_j})^2}}$$

After we calculate the weight of (Item 1, Item 2), (Item 2, Item 3), (Item 2, Item 4), we select K largest weight as the K most important neighbors of Item 2.

Lastly, we use the following formula to calculate the predicted rate of Item 2 rated by User 2.

$$P_{u, i} = \frac{\sum_{n\in N}r_{u, n}w_{i,n}}{\sum_{n\in N}|w_{i,n}|}$$

For Item CF with reviews, instead of selecting items with K largest weight as neighbors. We use the mean of cosine similarity of reviews written by the same users to select neighbors. For example, for Item 1 and Item 2, only User 1 and User 4 rated both items. In this case, we track the review embeddings written by User1 and User4 for both Item 1 and Item 2. Then, we calculate the cosine similarity of two reviews by the same user. In this case, Item 1 and Item 2 have two cosine similarity scores, one from User 1 and the other one from User 2. And we take the mean of these two cosine similarity scores as the similarity score of Item 1 and Item 2. After we calculate all similarity between pairs of Items, we select the pair of items with highest K similarity scores to do the same Item CF process.

\section{Experiment}
  \subsection{Selecting Samples}
  Firstly, the original data was downloaded from \url{https://www.yelp.com/dataset}, which is the official website of yelp. There are six data sets, and all of them are in the json format. For this project, we only need the data about business, users, and reviews. The size of the reviews data set is about 6.5 GB, and there are more than 965 million words in reviews. If each word is embedding a 300 dimension vector, the total size of only embedding words will be over 700 GB, which is unrealistic for NLP processing. Therefore, we decided to narrow down the restaurant location to a certain region and take 125k samples. We checked the number of restaurants in each state from business data set, and we found that the number of restaurants in Massachusetts is the largest (more than 30k) compared to other states, so we decided to focus on the restaurants only from this state. It will be helpful for the item-based filtering algorithm and our edited version algorithm to work well, and to see their difference if the users in sample data rating more restaurants, otherwise there is not enough review information to be analyzed. After trying different thresholds, we filter out users whose number of reviews is at least 35 in MA and the restaurants which have at least 150 reviews. Also, each review has at least 20 words after data cleaning. There are about 128k reviews that satisfy those conditions and 125k of them are randomly selected. With the data set ready, we decided to split it into two data set: one from training, and the other one for testing. The split ration between train data set and test data set is 4 : 1.
  
  \subsection{Preprocessing}
  A significant challenge for information extraction is that the vocabulary size is considerably large. We observed that people intend to use informal language while writing reviews, which involves using a massive amount of non-alphabetic characters, or misspelled words. Therefore, we decided to do preprocessing over the context by changing all the alphabetic characters to lowercase, restoring the abbreviations with non-alphabetic characters, and removing all the non-alphabetic characters and common stop words in English by using "ntlk" library. After the first round of preprocessing, we, however, found that the number of the unique words is 84,511, which is too large to train a word2vec model and do the embedding.
  
  \subsection{Misspelled word correction}
  Next, we decided to misspelled word correction by correcting or removing misspelled words to reduce vocabulary size. We observed that one of the most common spelling error types is that people like to repeat some characters consecutively in the same word to express a tone of emphasis, such as "goooodddd", "beauuutifffuuuul", "wooooonderrrrrfullll", etc. We also found that in nearly all of the English words, at most two consecutive duplications of the same character in a word are allowed, and the length of each duplication is at most two. With this observation, we first removed the excessive consecutive duplication of a character in a word.
  
  The removal of excessive consecutive duplication reduced the Damerau-Levenshitein edit distance of the most misspelled words, instead of correcting them. Meanwhile, the big-O time complexity of this process is linear regarding to the length of each word. Considering the size of the total sample, so it is very time consuming to apply it to all the words. Therefore, we used the "pyspellchecker" library from python to do the spelling error detection first, which will check if this word is in the provided dictionary of correct English words. If so, then we do not need to do anything about it. Otherwise, we will apply this algorithm to the misspelled words and get it ready for the next processing.
  
  Beside of providing spelling error correction, the "pyspellchecker" can provide a suggested correction for the misspelled word. This function is very limited. It cannot detect and provide suggested corrections for cognitive errors. Also, it can only provide a suggestion for the misspelled word with Damerau-Levenshitein edit distance at most two. For the misspelled words that cannot be corrected, we had to remove it to reduce the vocabulary size and avoid over fitting. Therefore, with the above algorithm and correction function from "pyspellchecker", we reduced the vocabulary size to 73,061.
  
  \subsection{Lemmatization}
  Meanwhile, we also tried to perform the lemmatization over the sample, which is grouping the inflected forms of a word, changing them into the base form, and analyzing them as a single form. We used the "nltk" library to perform this process. With the lemmatization, the total vocabulary size went down to 67,233. However, lemmatization can be lossy. The word with different forms can deliver different meanings depending on the context it is in. Also, people like to express their feelings with the different forms of a word, and lemmatization can make this semantic information lost. 
  
  \subsection {Word Embedding}
  Out first approach for work embedding is to use the word2vec model, and mainly work with the "gensim" library from python. We selected two pre-trained word2vec model provided by "gensim": "word2vec-google-news-300" and "glove-wiki-gigaword-300". Also, the "gensim" library allowed us to train our own word2vec model based on our data set. Because the "word2vec-google-news-300" model only has the option of 300 embedding vectors, we set the vector size of our own model to 300 for consistency. 
  
  We also use Bert to encoding sentences directly. We import sentence_transformers packages in python. In this package, we import 'bert-base-nli-mean-tokens' pre-trained model to avoid too long training time. For the pretrained Bert, it keeps a max 128 sentence length and does a mean pooling to generate a 768 dimension output sentence encoding embedding. By using pretrained Bert model, we successfully reduce the training time from days to less than 30 minutes. 
  
  
  \subsection{Computing Similarity}
  There are three word embedding models in this project: word2vec-google-news-300 model (Google), training a word2vec model (Own W2V), glove-wiki-gigaword-300 model (GloVe); and one sentence embedding model: Bert model (Bert). The way of computing similarity between two reviews are different depending on the chosen embedding method. For Bert model, the similarity between two review sentences is simply the cosine similarity of their sentence embedding vectors. For the first three models, we first embed all words in every review to vectors. Then we applied the FaceBook InferSent model, which is a BiLSTM with Max-Pooling, to extract the most important 512 features of each sentence as a sentence vector. Finally, we calculate cosine similarity among embedding vectors as the similarity between two sentences.
  
  \subsection{Item CF}
  We calculated the RMSE of original Item CF as the gold standard score, and compared the RMSE scores of Item CF with reviews to see how much we can improve. For the original item CF, the model is basically the same as described above. Meanwhile, we compared the RMSE scores of three selecting neighbors methods. The first one is selecting top K neighbors with K = 5,10,15, 20, 25, and 30. The second one is keeping all weights as neighbors to calculate RMSE score. The third one is only deleting all pairs with negative weights. In this case, we only use non-negative weights to predict. After comparing the RMSE score of all three methods,  the best RMSE is 1.082271
  
  For Item CF with reviews, we first calculated the neighbors of each test case. To reduce the running time, we limited the candidates to restaurants users rated. We reduced the running time from over 6 hours to less than 1 hour. Then, we selected restaurants with top 10 highest similarity scores. For all neighbors for each test user, we calculated the weights of them with the test restaurant, and then predicted the according rating. The whole process was repeated eight times. For each embedding encoding method (Bert, Google, Glove, Own W2V), we calculated the RMSE of two cases: one for reviews with lemmatization and the other for reviews without lemmatization. All hyperparameters used in Item CF remain same. By controlling variables, we compared the performance of different embedding encoding methods, and differentiated whether lemmatization can emphasize the key features of reviews.
\section{Result \& Discussion}

According to the result RMSEs in table 2, we can see that Bert catches the most important features and generates the most relevant similar neighbors for Item CF by comparing the result RMSE of each method. Lemmatization helps Bert, Glove, and Own W2V to capture more important features so as to reduce RMSE scores of Item CF. Only for Google embedding method, encoding without lemmatization helps to better predict ratings of Item CF. We think that this is due to the lost of information of some helpful features in the reviews from different pretrained vocabulary. For our trained models such as Bert and Word2Vec, we can see lemmatization helps to customize more on our own training reviews.

\begin{tabular}{ |p{2.5cm}||p{2.5cm}|p{1.3cm}| } 
\hline
Embedding & Item CF with Review & Original Item CF \\
\hline
Bert with lemma & 0.931494 & \multirow{8}{1.3cm}{1.082271}\\ 
\cline{1-2}
Bert without lemma & 0.939019 & \\
\cline{1-2}
Glove with lemma & 0.947504 &\\
\cline{1-2}
Glove without lemma & 0.947704 &\\
\cline{1-2}
Google with lemma & 0.948165 &\\
\cline{1-2}
Google without lemma & 0.945793 &\\
\cline{1-2}
Own W2V 300d with lemma & 0.947019 &\\
\cline{1-2}
Own W2V 300d without lemma & 0.948520 &\\
\cline{1-2}
Own W2V 200d with lemma & 0.948401 &\\
\cline{1-2}
Own W2V 200d without lemma & 0.949589&\\
\hline
\end{tabular}
\captionof{table}{Result table.}

The main problem remained is our current Item CF cannot properly predict extreme ratings such as 1 and 5. In most cases, our Item CF with reviews predicts those pairs as 3.7 - 4.1. By eliminating all ratings 5 in the test case can dramatically reduce RMSE to 0.79. We think this is due to the limitation of our computational power so that we cannot involve more data in our running process. For the candidates of Item CF with review, we only select candidates from user rated business from the 100,000 training reviews  instead of all business. Even with the reduced size of our data, we currently still need to train our models for hours. With more powerful computation resource, We think that we can involve more samples in training and more neighbors for Item CF to predict which can continuously improve the performance of Item CF with reviews. Also, We can try to combine the result of Original Item CF and that of the Item CF with reviews with assigning different weight to the result of each model. For example, since the original Item CF predicts extreme ratings better than our current Item CF with reviews, we may use some weighting and switching methods.

\section{Conclusion}
During the preprcoessing stage, we reduced the vocabulary size by 11,450. However, this does not match our original expectation. One of the reasons is that the default dictionary of ”pyspellchecker” library is not rich enough for slangs. Our next step will try to use the urban dictionary API, which can provide much richer dictionary for slangs. Also, we noticed that our own word2vec model only calculated 19,417 words, which is much less that our vocabulary. One of plausible explanations is that the gensim will skip some word with low frequency.

Regarding about the resulst from different embedding methods, we found that BERT embedding only has insignificant advantage comparing with the word2vec model. Other than the lack of training, we believe that the word2vec model was affected by the preprocess procedure over the data set. With better spelleing correction techniques, we believe that different between BERT and word2vec can be more significant.

In the item CF with review, we extracted semantic information from users to restaurants, which provided richer information than the rating system in the original item CF system. Due to the limitation in computation resource, we could not train our system with more epochs, and the results from RSME indicated that our model has not converged enough. If we could have the access to more powerful resource, we are confident that the performance of our system can be improved. In conclusion, comparing with the original item CF, our hybrid system with semantic analysis over the user's comment can extract much information about user's preference, and provide better recommendations.

\section{Related Links}
Link to our demo video on YouTube \\
\url{https://www.youtube.com/watch?v=rjlxWL2sHas}

Link to our GitHub repository\\
\url{https://github.com/bigshawne/CSCI_544_Final_Proj_Recommendation_System_Based_On_NLP}

\bibliographystyle{acl}
\bibliography{acl2015}

\end{document}